\documentclass{article}

\usepackage{PRIMEarxiv}

\usepackage[utf8]{inputenc} 
\usepackage[T1]{fontenc}    
\usepackage{hyperref}       
\usepackage{url}            
\usepackage{booktabs}       
\usepackage{amsfonts}       
\usepackage{nicefrac}       
\usepackage{microtype}      
\usepackage{lipsum}
\usepackage{fancyhdr}       
\usepackage{graphicx}       
\graphicspath{{media/}}     

\title{QUARK MATTER IN THE NJL MODEL WITH A VECTOR INTERACTION
AND THE STRUCTURE OF HYBRID STARS
}

\author{
  G.B. Alaverdyan \\
    Yerevan State University \\
  Armenia\\
  \ galaverdyan@ysu.am} 

\begin{document}
\maketitle

\begin{abstract}

{The properties of hadron-quark hybrid stars are studied when the quark phase is described in terms of a local SU(3) Nambu-Jona-Lasinio (NJL) model taking into account the contribution of the vector and axial-vector interaction between the quarks, and the hadronic phase, in the relativistic mean field (RMF) model. For different values of the vector coupling constant $G_V$, the equations of state of the quark matter are calculated and the parameters of the hadron-quark phase transition are determined under the assumption that the phase transition takes place in accordance with Maxwell’s construction. It is shown that for a larger vector coupling constant, the equation of state of the quark matter will be “stiffer” and the coexistence pressure $P_0$ of the phases will be greater. Using the resulting hybrid equations of state, the TOV equations are integrated numerically and the mass and radius of the compact star are determined for different values of the central pressure $P_c$. It is shown that when $G_V$ is larger, the maximum mass of the compact star will be larger and thereby, the radius of the configuration with maximum mass will be smaller. Questions of the stability of hybrid stars are also discussed. It is shown that in terms of the model examined here, for all values of the vector coupling constant, a hybrid star with an infinitely small quark core is stable. These results are compared with recent measurements of the mass and radius of the pulsars PSR J0030+0451 and PSR J0740+6620, carried out at the International Space Station with the NASA's Neutron star Interior Composition Explorer (NICER) X-ray telescope. A comparison of the theoretical results with observational data does not exclude the possibility of quark deconfinement in the interiors of compact stars.}

\end{abstract}


\section{Introduction}
Studies of the properties and composition of compact stars are an important area of modern physics. It is known that matter in these kind of celestial objects has a low temperature and an extremely high density in the central part of the object. To describe  the structure of a compact star it is necessary to know the equation of state of strong interacting matter at the baryon number densities $n_B\in[0-10n_0]$, where $n_0=0.16$ fm$^{-3}$ is the saturation density of nuclear matter. The matter in the central part of a compact star is essentially a system of particles described by quantum chromodynamics (QCD). At the extremely high densities inside a compact star various exotic structures may appear, such as hyperonic matter, pion condensate, kaon condensate, a quark phase, a color superconducting phase.

Over the last few decades, many papers have dealt with the study of the thermodynamic properties of quark matter, obtaining the corresponding equation of state, and the application of these results to the study of the properties of compact stars. Work in this area has become more intense, especially after the discovery of the two pulsars PSR J1614-2230 \cite{1} and PSR J0348+0432 \cite{2} with a mass on the order of twice that of the sun.

Exact measurements of the mass of compact stars are a good way of obtaining limits on the equation of state of superdense matter inside a star. Simultaneous measurement of the mass and radius of a neutron star undoubtedly opens up new possibilities of obtaining stricter limitations on the equation of state of superdense matter. 

Recent measurements with the NICER X-ray telescope installed on the International Space Station have made it possible simultaneously to measure both the mass and radius of PSR J0030+0451. The following values have been obtained for this pulsar: $M=1.44^{+0.15}_{-0.14}~ M_\odot$, $R=13.02^{+1.24}_{-1.06}$ km \cite{3}. Observations of the pulsar PSR J0740+6620 and an analysis of the data in the framework of this same program have made it possible to determine the mass and radius: $M=2.08^{+0.07}_{-0.07}~ M_\odot$, $R=13.7^{+2.6}_{-1.5}$ km \cite{4,5}. We note that this pulsar is the most massive neutron star known at this time.

In many studies of the quark phase inside hybrid stars (e.g., Refs.\cite{{6},{7},{8},{9},{10},{11},{12},{13},{14},{15},{16},{17}}), the MIT quark bag phenomenological model  \cite{18} is used. Recetly, the Nambu--Jona-Lasinio (NJL) model \cite{19,20} has very often been used for describing quark matter. This model successfully reproduces many features of quantum chromodynamics \cite{{21},{22},{23},{24},{25}}. By combining different modifications of the quark NJL model with various models describing hadronic matter, a number of authors have constructed a quark-hadron hybrid equation of state (e.g., Refs.\cite{{26},{27},{28},{29},{30}}).

In our paper \cite{31} we studied the structure of hybrid stars using an equation of state with a quark phase transition at a constant pressure, when the hadronic phase is described in a relativistic mean field (RMF) model \cite{32,33}, while the quark phase is described by a local SU(3) NJL model, where the terms of the vector  and axial-vector channels of the quark interaction were neglected. Recently, we have studied the thermodynamic characteristics of quark matter in terms of local NJL model, in which the terms of the vector and axial-vector interactions were also taken into account \cite{34}.  

The purpose of this paper is to obtain a hybrid equation of state in terms of the local NJL model with the vector and axial-vector interaction channels taken into account for the quarks at different values of the vector interaction coupling $G_V$ and to study the influence of this type of interaction on the structure of a compact star. The hadronic phase is described using an expanded RMF model where, besides the fields of the exchanged $\sigma$-, $\omega$-, and $\rho$- mesons, the scalar-isovector $\delta$-meson \cite{11} also taken into account. 

In constructing the hybrid equation of state it was assumed that the surface tension between the quark and hadronic phases is so strong that the formation of a mixed phase is energetically unfavorable and the phase transition takes place at a constant pressure in accordance with a Maxwell construction. This assumption is also supported by the conclusion of a recent paper \cite{35} to the effect that repulsive vector interactions strongly enhance the surface tension of the quark matter.

In this paper we shall use the "natural" system of units in which  $\hbar=c=1$.

\section{The hadronic phase in the RMF model with the contribution of the scalar-isovector $\delta$-meson field taken into account}
In this paper we have assumed that in the supernuclear density region electrically neutral hadronic matter in $\beta$ equilibrium consists of protons $p$, neutrons $n$, and electrons $e$. For the thermodynamic description of hadronic matter, the relativistic mean field (RMF) theory based on a quantum-field model of meson exchange was used. To describe the strong interaction between nucleons the channels corresponding to exchange mesons with different transformation properties are taken into account in isotopic and ordinary space: the isoscalar-scalar $\sigma$-meson, isoscalar-vector $\omega$-meson, isovector-scalar $\delta$-meson, and isovector-vector $\rho$-meson.

The density of the relativistic Lagrangian of the hadronic matter consisting of protons, neutrons, and electrons in terms of the RMF model is given by \cite{11}

\begin{eqnarray}
\label{eq1} 
{ \cal L}_{RMF}=\bar {\psi} _{N}\left[ \gamma ^{\mu
}\left(i\partial _{\mu }-g_{\omega }\omega _{\mu }(x)-\frac{1}{2}g_{\rho }%
\overrightarrow{\tau }_N\overrightarrow{\rho }_{\mu
}(x)\right) -\left( m_{N}-g_{\sigma }\sigma (x)-g_{\delta
}\overrightarrow{\tau }_{N}\overrightarrow{\delta }(x)\right)
\right] \psi _{N} \nonumber \\
+\frac{1}{2}\left( \partial _{\mu }\sigma (x)\partial ^{\mu
}\sigma (x)-m_{\sigma }\sigma (x)^{2}\right) -\frac{b}{3}m_N \left(g_\sigma \sigma(x)\right)^3-\frac{c}{4} \left(g_\sigma \sigma(x)\right)^4 \nonumber \\
+\frac{1}{2}m_{\omega }^{2}\omega ^{\mu }(x)\omega _{\mu
}(x)-\frac{1}{4}\Omega _{\mu \nu}(x)\Omega ^{\mu \nu }(x) 
+\frac{1}{2}\left( \partial _{\mu }\overrightarrow{\delta
}(x)\partial
^{^{\mu }}\overrightarrow{\delta }(x)-m_{\delta }^{2}\overrightarrow{%
\delta }(x)^{2}\right) \\ 
+\frac{1}{2}m_{\rho }^{2}\overrightarrow{\rho }^{\mu }(x)%
\overrightarrow{\rho }_{\mu }(x)-\frac{1}{4}\Re_{\mu \nu
}(x)\Re^{\mu \nu}\left( x\right)+\bar {\psi} _{e}\left( i\gamma ^{\mu
}\partial _{\mu }-m_e \right)\psi _{e} \nonumber
\end{eqnarray}

Here   $\psi _{N} =
\left({{\begin{array}{*{20}c} {\psi _{p}}  \hfill \\ {\psi _{n}}
\hfill \\\end{array}} } \right)$ is the isospin doublet of the nucleon bispinor, 
$\psi_e$ is the electron wave function,  $\vec {\tau}_N$ are
the isospin $2\times2$ Pauli matrices, $\sigma(x)$, $\omega
_{\mu }(x)$, $\overrightarrow{\delta }(x)$, and
$\overrightarrow{\rho }_{\mu }(x)$ are the fields of the exchange mesons at the space-time point $x=x_{\mu}=(t,x,y,z)$, $m_N$, $m_e$, $m_\sigma$, $m_\omega$, $m_\delta$, and $m_\rho$ are the masses of the free particles, $\Omega_{\mu \nu }(x)$ and $\Re_{\mu \nu
}(x)$ are the antisymmetric tensors of the vector fields $\omega
_{\mu }(x)$ and $\overrightarrow{\rho }_{\mu }(x)$, respectively:

\begin{equation}
\label{eq2} \Omega _{\mu \nu}  \left( {x} \right) = \partial
_{\mu}  \omega _{\nu} \left( {x} \right) - \partial _{\nu}  \omega
_{\mu}  \left( {x} \right),\quad \;\Re _{\mu \nu}  \left( {x}
\right) = \partial _{\mu}  \rho _{\nu}  \left( {x} \right) -
\partial _{\nu}  \rho _{\mu}  \left( {x} \right).
\end{equation}

In Eq. (1), $g_\sigma$, $g_\omega$, $g_\delta$, and $g_\rho$ denote the coupling constants of a nucleon with the corresponding exchange meson,
$b$ and $c$ the constants characterizing the contribution of the nonlinear part of the potential of the $\sigma$-field owing to the self-interaction of the $\sigma$-mesons \cite{36}.

In the RMF approximation the meson fields are replaced by effective average fields. The energy density and pressure of the hadronic matter in this approximation take the form (for details, see Ref. 11)

\begin{eqnarray}{}\label{eq3}
\varepsilon_{HP} =
\frac{{1}}{{\pi^{2}}}\int\limits_{0}^{k_{F} (n_B) \left( {1 - \alpha} \right)^{1/3}} {dk~k^{2}\sqrt {k^{2} + \left(
{m_{N} - \sigma - \delta}  \right)^{\,2}}}+\frac{b}{3}m_N \sigma^3+\frac{c}{4} \sigma^4  \nonumber\\ 
+\frac{{1}}{{\pi ^{2}}}\int\limits_{0}^{k_{F\,} \left( {n_B}
\right)\left( {1 + \alpha}  \right)^{1/3}} {dk~k^{2}\sqrt {k^{2} + \left(
{m_{N} - \sigma + \delta}  \right)^{\,2}}} + \frac{{1}}{{2}}\left(
{\,\frac{{\sigma ^{\,2}}}{{a_{\sigma} } } + \,\frac{{\omega
^{2}}}{{a_{\omega} } } + \frac{{\delta ^{\,2}}}{{a_{\delta} } } +
\frac{{\rho ^{\,2}}}{{a_{\rho} } }}\right)\\
+\frac{{1}}{{\pi^{2}}}\int\limits_{0}^{\sqrt{\mu_e^2-m_e^2}} {dk~k^{2}\sqrt {k^{2} + m_e^2}}\nonumber~~~ ,
\end{eqnarray}

\begin{eqnarray}{}\label{eq4}
P_{HP} =
\frac{{1}}{{\pi^{2}}}\int\limits_{0}^{k_{F} (n_B) \left( {1 - \alpha} \right)^{1/3}} {dk~k^{2}\left(\sqrt {k_F(n_B)^{2}(1-\alpha )^{2/3} + \left({m_{N} - \sigma - \delta}  \right)^{2}}-\sqrt {k^{2} + \left({m_{N} - \sigma - \delta}  \right)^{\,2}}\right)}\nonumber\\ +\frac{{1}}{{\pi^{2}}}\int\limits_{0}^{k_{F} (n_B) \left( {1 + \alpha} \right)^{1/3}} {dk~k^{2}\left(\sqrt {k_F(n_B)^{2}(1+\alpha )^{2/3} + \left({m_{N} - \sigma + \delta}  \right)^{2}}-\sqrt {k^{2} + \left({m_{N} - \sigma + \delta}  \right)^{\,2}}\right)}\nonumber\\
-\frac{b}{3}m_N \sigma^3-\frac{c}{4} \sigma^4 + \frac{{1}}{{2}}\left(
{-\frac{{\sigma ^{\,2}}}{{a_{\sigma} } } + \frac{{\omega
^{2}}}{{a_{\omega} } } - \frac{{\delta ^{\,2}}}{{a_{\delta} } } +
\frac{{\rho ^{\,2}}}{{a_{\rho} } }}\right)\\
-\frac{{1}}{{\pi^{2}}}\int\limits_{0}^{\sqrt{\mu_e^2-m_e^2}} {dk~k^{2}\sqrt {k^{2} + m_e^2}}+\frac{1}{3\pi^2}\mu_e\left( \mu_e^2-m_e^2 \right)^{3/2}\nonumber~~~ ,
\end{eqnarray}

where $n_B$ is the density of the baryonic charge of the hadronic matter, $\alpha=(n_n-n_p)/n_B$ is the asymmetry parameter, $\mu_e$ is the chemical potential of an electron, and $k_F(n_B)=\left(3\pi^2n_B/2\right)^{1/3}$. The redefined mean meson-fields $\sigma$, $\omega$, $\delta$, and $\rho$ in Eqs. (3) and (4) are expressed in terms of the average fields of the corresponding mesons in the following way:

\begin{equation}
\label{eq5} 
\sigma \equiv g_{\sigma}  \langle {\sigma}(x) \rangle  \,,\quad
\omega \equiv g_{\omega}  \langle {\omega}(x) \rangle \,,\quad 
\delta \equiv g_{\delta}  \langle {\delta}(x) \rangle \,,\quad 
\rho \equiv g_{\rho}  \langle {\rho}(x) \rangle \,.
\end{equation}

The model parameters $a_\sigma$, $a_\omega$, $a_\delta$, and $a_\rho$ are expressed in terms of the coupling constants and masses of
the mesons as:

\begin{equation}
\label{eq6} 
a_{\sigma}\equiv (g_{\sigma}/m_{\sigma})^{2}, \quad
a_{\omega}\equiv (g_{\omega}/m_{\omega})^{2}, \quad
a_{\delta}\equiv (g_{\delta}/m_{\delta})^{2}, \quad
a_{\rho}\equiv (g_{\rho}/m_{\rho})^{2}\,
\end{equation}

The average meson-fields satisfy the equations

\begin{equation}
\label{eq7} 
\sigma = a_{\sigma}  \left( {n_{sp} \left(
{n_B,\alpha}  \right) + n_{sn} \left( {n_B,\alpha}  \right)\, -
b\,m_{N} \sigma ^{2} - c\,\sigma ^{3}} \right),
\end{equation}
\begin{equation}
\label{eq8} 
\omega = a_{\omega}  n_B,
\end{equation}
\begin{equation}
\label{eq9} 
\delta = a_{\delta}  \left( {n_{sp} \left({n_B,\alpha}  \right) - n_{sn} \left( {n_B,\alpha}  \right)\,}
\right),
\end{equation}
\begin{equation}
\label{eq10} \rho = - \frac {1}{2} a_{\rho}n_B\,\alpha \,.
\end{equation}

The scalar densities of the protons and neutrons $n_{sp} ( {n_B,\alpha})$  and $n_{sn}( {n_B,\alpha})$ are defined by

\begin{equation}
\label{eq11} 
n_{s\,p} \left( {n_B,\alpha}  \right) = \frac{{1}}{{\pi
^{2}}}\int\limits_{0}^{k_{F\,} \left( {n_B} \right)\left( {1 -
\alpha} \right)^{1/3}} dk\, k^{2} {\frac{{m_N-\sigma-\delta}}{{\sqrt {k^{2} + \left(m_N-\sigma-\delta\right)^2}} }} \;\; \quad ,
\end{equation}

\begin{equation}
\label{eq12} 
n_{s\,n} \left( {n_B,\alpha}  \right) = \frac{{1}}{{\pi
^{2}}}\int\limits_{0}^{k_{F\,} \left( {n_B} \right)\left( {1 +
\alpha} \right)^{1/3}} dk\, k^{2} {\frac{{m_N-\sigma+\delta}}{{\sqrt {k^{2} + \left(m_N - \sigma+\delta\right)^2}} }} \;\; \quad ,
\end{equation}

The chemical potentials of a proton, neutron, and electron are expressed in terms of the baryonic charge density $n_B$, the asymmetry parameter $\alpha$, and the meson mean-fields

\begin{eqnarray}
\label{eq13} 
\mu_p(n_B,\alpha)=\sqrt {k_{F} \left( {n_B}
\right)^{2}\left( {1 - \alpha} \right)^{2/3} + \left( {m_{N} -
\sigma - \delta}
\right)^{\,2}} + \omega + \frac{1}{2}\rho,\nonumber\\
\mu_{n}(n_B,\alpha)=\sqrt {k_{F} \left( {n_B} \right)^{2}\left( {1 +
\alpha} \right)^{2/3} + \left( {m_{N} - \sigma + \delta}
\right)^{\,2}}+ \omega - \frac{1}{2}\rho,\\
\mu_{e}(n_B,\alpha)=\sqrt {\left( \frac{3}{2}\pi^2n_B(1-\alpha)\right)^{2/3}+{m_e}^2}.\nonumber
\end{eqnarray}

For the matter in a neutron star consisting of protons, neutrons, and electrons, the condition for $\beta$-equilibrium after emergence from the system of all neutrinos will have the form
\begin{equation}
\label{eq14} 
\mu_n(n_B,\alpha)=\mu_p(n_B,\alpha)+\mu_e(n_B,\alpha) .
\end{equation}

The system of Eqs. (7)-(10) and (14) for a specified value of the baryonic density $n_B$ can be solved to find the asymmetry parameter $\alpha$ and the mean meson-fields $\sigma$, $\omega$, $\delta$ and $\rho$. Knowledge of these quantities makes it possible to calculate the energy density and pressure for a specified value of the baryonic density $n_B$. Thus, carrying out the above algorithm for the numerical calculation lets us have an equation of state for cold hadronic matter in the parametric form $\{ \varepsilon_{HP}(n_B); P_{HP}(n_B) \}$.

In carrying out these numerical calculations we have used the values for the parameters of the RMF model given in Ref. 11: 
$a_\sigma=9.154$  fm$^2$, $a_\omega=4.828$ fm$^2$, $a_\delta=2.5$ fm$^2$,  $a_\rho=13.621$ fm$^2$, $b=1.654~10^{-2}$  fm$^{-1}$, and $c=1.319~10^{-2}$  fm$^{-2}$.
 
\section{Quark phase in the NJL model with the vector interaction taken into account}

For describing the properties of the quark phase in a neutron star we used a local SU(3) NJL model in which, besides the scalar channel for the interaction between quarks, a vector interaction channel is also taken into account.
The Lagrangian density for the system consisting of the $u$, $d$, and $s$ quarks and electrons in terms of this model has the form:

\begin{equation}
\label{eq15} 
{\cal L}_{RMF}= {\cal L}_{F}+{{\cal L}_F}^{(e)}+{\cal L}_{S}+{\cal L}_{D}+{\cal L}_{V}.
\end{equation}

Here the first term ${\cal L}_{F}$ is the density of the Dirac Lagrangian for the fields of the free quarks:

\begin{equation}
\label{eq16} 
{\cal L}_{F}=\overline{\psi}\left(i\gamma ^{\mu }
\partial _{\mu }-\hat{m}_0\right)\psi,
\end{equation}

where $\psi$ is the spinor field of the ${\psi_f}^c$c quarks with three flavors $f = u,~ d,~ s$ and three colors $c = r,~ g,~ b$, while $\hat{m}_0$ is
the diagonal matrix of the current masses of the quarks ${\hat{m}_0}=$ diag $(m_{0u},~m_{0d},~m_{0s})$.

The second term ${{\cal L}_F}^{(e)}$ is the density of the Dirac Lagrangian for the system of free electrons:

\begin{equation}
\label{eq17} 
{{\cal L}_F}^{(e)}=\overline{\psi}_e\left(i\gamma ^{\mu }
\partial _{\mu }-m_e\right)\psi_e,
\end{equation}

${{\cal L}_S}$ corresponds to the four-quark chirally-symmetric interaction of a scalar and pseudoscalar type with a coupling constant $G_S$:

\begin{equation}
\label{eq18} 
{\cal L}_S=G_S \sum_{a=0}^{8}\left[(\overline{\psi}\lambda_a\psi)^2+(\overline{\psi}i\gamma_5\lambda_a\psi)^2\right],
\end{equation}

where $\lambda_a$ ($a = 1, 2, ..., 8$) are the Gell-Mann matrices and simultaneously the generators of the SU(3) group in flavor space, while $\lambda_0=\sqrt{2/3}\hat I$ ( $\hat I$ is the unit 3 $\times 3$ matrix).

The term ${\cal L}_D$ in the Lagrangian (15) corresponds to the Kobayashi-Maskawa-’t Hooft six-quark interaction \cite{37} and has the form

\begin{equation}
\label{eq19} 
{\cal L}_D=-K \left\{ det_f \left(\overline{\psi}(1+\gamma_5)\psi\right)+det_f \left(\overline{\psi}(1-\gamma_5)\psi\right) \right\}.
\end{equation}

Introducing this kind of interaction term in the Lagrangian has made it possible to reproduce the value of the masses of the pseudoscalar isosinglet mesons $\eta'(958)$ and $\eta(547)$ in the NJL model.

The last term ${\cal L}_V$ in the Lagrangian density (15) is the vector and axial-vector interaction among the quarks with a vector coupling constant $G_V$:

\begin{equation}
\label{eq20} 
{\cal L}_V=-G_V \sum_{a=0}^{8}\left[(\overline{\psi}\gamma_{\mu}\lambda_a\psi)^2+(\overline{\psi}i\gamma_{\mu}\gamma_5\lambda_a\psi)^2\right],
\end{equation}

By using the mean field approximation from the Lagrangian (15) for the energy density and pressure of the
quark phase one can obtain \cite{34}

\begin{eqnarray}
\label{eq20} 
\varepsilon_{QP}=\frac{3}{\pi^2}\sum_{f=u,d,s}\left[~{\int_0^\Lambda{dk~k^2\sqrt{k^2+{M_{f0}}^2}}
-\int_{(\pi^2n_f)^{1/3}}^\Lambda{dk~k^2\sqrt{k^2+{M_{f}}^2}}}~\right] \nonumber \\
+2G_S({\sigma_u}^2+{\sigma_d}^2+{\sigma_s}^2
-{\sigma_{u0}}^2-{\sigma_{d0}}^2-{\sigma_{s0}}^2)-4K(\sigma_u~\sigma_d~\sigma_s-\sigma_{u0}~\sigma_{d0}~\sigma_{s0})\\
+2G_V({n_u}^2+{n_d}^2+{n_s}^2)+\frac{1}{\pi^2}\int_0^{(3\pi^2n_e)^{1/3}}{dk~k^2\sqrt{k^2+{m_e}^2}}~~,\nonumber
\end{eqnarray}

\begin{eqnarray}
\label{eq21} 
P_{QP}=-\frac{3}{\pi^2}\sum_{f=u,d,s}\left[~{\int_0^\Lambda{dk~k^2\sqrt{k^2+{M_{f0}}^2}}
-\int_{(\pi^2n_f)^{1/3}}^\Lambda{dk~k^2\sqrt{k^2+{M_{f}}^2}}}~\right] \nonumber \\
+\sum_{f=u,d,s}n_f\sqrt{(\pi^2n_f)^{2/3}+{M_f}^2}-2G_S({\sigma_u}^2+{\sigma_d}^2+{\sigma_s}^2
-{\sigma_{u0}}^2-{\sigma_{d0}}^2-{\sigma_{s0}}^2)\\
+4K(\sigma_u~\sigma_d~\sigma_s-\sigma_{u0}~\sigma_{d0}~\sigma_{s0})+2G_V({n_u}^2+{n_d}^2+{n_s}^2)\nonumber \\
-\frac{1}{\pi^2}\int_0^{(3\pi^2n_e)^{1/3}}{dk~k^2\sqrt{k^2+{m_e}^2}+n_e\sqrt{(3\pi^2n_e)^{2/3}+{m_e}^2}}~~.\nonumber
\end{eqnarray}

Here $M_f$ is the constituent or dynamic mass of a quark with flavor $f$ in this state, and $M_{f0}$ is the constituent
mass for $n_u=n_d=n_s=0$ . $\sigma_f$ and $\sigma_{f0}$ ($f = u,~ d, ~s$) denote the quark condensates $\langle \bar{\psi}_f$ $\psi_f\rangle$, respectively, in this state and in the vacuum ( $n_u = n_d = n_s = 0$), which are defined as

\begin{equation}
\label{eq23} 
\sigma_f=-\frac{3M_f}{\pi^2}\int_{(\pi^2n_f)^{1/3}}^\Lambda
dk~\frac{k^2}{\sqrt{k^2+{M_f}^2}}~,~~~\sigma_{f0}=-\frac{3M_{f0}}{\pi^2}\int_{0}^\Lambda
dk~\frac{k^2}{\sqrt{k^2+{M_{f0}}^2}}~.
\end{equation}

Since the model examined here is nonrenormalizable, there is a need for an ultraviolet cutoff in the diverging integrals. $\Lambda$ is the ultraviolet cutoff parameter in momentum space.

The chemical potentials of the particles in the quark phase are determined by the expressions

\begin{eqnarray}
\label{eq24} 
\mu_f=\sqrt{(\pi^2n_f)^{2/3}+{M_f}^2}+4\,G_V\,n_f,~~(f=u,d,s),
~~~~~\mu_e=\sqrt{(3\pi^2n_e)^{2/3}+{m_e}^2}.
\end{eqnarray}

In the NJL model the constituent masses of the quarks satisfy the “gap” equations
\begin{eqnarray}
\label{eq25} 
 M_u=m_{0u}-4\,G_S\,\sigma_u+2\,K\sigma_d~\sigma_s, \nonumber \\
 M_d=m_{0d}-4\,G_S\,\sigma_d+2\,K\sigma_s~\sigma_u, \\
 M_s=m_{0s}-4\,G_S\,\sigma_s+2\,K\sigma_u~\sigma_d.  \nonumber
\end{eqnarray}

Assuming that all the neutrinos have been able to leave the system, the condition for $\beta$-equilibrium can written
in the form
\begin{equation}
\label{eq26} 
\mu_d=\mu_u+\mu_e\,;~~~\mu_s=\mu_d\,.
\end{equation}

For electrically neutral quark matter we shall have
\begin{equation}
\label{eq27} 
\frac{2}{3}n_u-\frac{1}{3}n_d-\frac{1}{3}n_s-n_e=0\,.
\end{equation}
The baryonic charge density is defined by
\begin{equation}
\label{eq28} 
n_B=\frac{1}{3}(n_u+n_d+n_s)\,.
\end{equation}

On numerically solving the system of equations (25)-(28) for a specified value of the baryonic density $n_B$ it is possible to find the constituent masses of the quarks $M_u$, $M_d$, and $M_s$ as well as the particle concentrations $n_u$, $n_d$, $n_s$, and $n_e$. Equations (21) and (22) will then represent the equation of state of the quark phase in the parametric form $\{ \varepsilon_{QP}(n_B); P_{QP}(n_B) \}$ .

In carrying out the numerical calculations we have used the values for the parameters of the NJL model given in Ref. 23: $m_u = m_d = 5.5$ MeV, $m_s = 140.7$ MeV, $\Lambda=602.3$ MeV, $G_S=1.835/\Lambda^2$, and $K=12.36/\Lambda^5$. In Ref. 23 these
parameters of the NJL model were obtained by fitting from a reproduction of the values of the masses of the pseudoscalar $\pi$, $K$, and $\eta'$ mesons, as well as the decay constant $f_\pi$ of the pion.

The vector coupling constant $G_V$ is a free parameter in this paper. In order to study the effect of the vector interaction on both the parameters of the hadron-quark phase transition and on the properties and structure of a hybrid star, we have used several values of this constant: $G_V=0$,  $G_V=0.2\,G_S$, $G_V=0.4\,G_S$, and $G_V=0.6\,G_S$.

\section{The hadron-quark phase transition and the hybrid equation of state for cold $\beta$-equilibrium electrically neutral superdense matter}

To obtain the hybrid equation of state with a hadron-quark phase transition it is necessary to know, not only the equation of state of the hadronic matter and the quark matter individually, but also the type of phase transition between these phases. It has been shown \cite{38} that in the case of a hadron-quark phase transition, because there are two conserved quantities, the phase transition may take place with formation of a mixed phase. In the mixed phase the condition of global electrical neutrality is fulfilled, when the quark matter and the hadronic matter are separately electrically charged. With this sort of phase transition the energy density and density of the baryonic charge are continuous in character, as opposed to the usual phase transition of the first-order, where these quantities vary discontinuously, while the pressure then remains constant. For an ordinary phase transition with constant pressure, the parameters of the transition are determined by constructing a common tangent to the curves $P(1/n_B)$ of the individual phases (a Maxwell construction). In the case of a transition of this type each of the phases is separately electrically neutral. 

Which of the above scenarios for a hadron-quark phase transition takes place in reality depends on the amount of energy of electrostatic and surface natures required to form the different geometric structures in the mixed phase consisting of hadronic matter and quark matter. For sufficiently high values of the surface tension coefficient of the quark matter, formation of geometric structures in the mixed phase will be energetically unfavorable and the phase transition will take place in accordance with a Maxwell construction.

The unknowability of the surface tension coefficient of the quark matter makes it impossible to uniquely determine which of the two scenarios of the hadron-quark phase transition actually takes place. In this paper we assume that the phase transition of the deconfinement takes place in a Maxwell construction scenario, i.e., at a constant pressure $P = P_0$ and with a discontinuous change in the baryonic density from $n_{H0}$ to $n_{Q0}$.

The parameters of the phase transition can be found by solving the equations for the conditions of thermal,mechanical, and chemical equilibrium between the two phases. For cold matter, these equations have the form

\begin{equation}
\label{eq29} 
T_{HP}=T_{QP}=0,~~~P_{HP}=P_{QP}=P_0,~~~{\mu_B}^{HP}={\mu_B}^{QP}=\mu_{B0}\,,
\end{equation}
where ${\mu_B}^{HP}=\mu_n$ is the baryonic chemical potential in the hadronic phase, and ${\mu_B}^{QP}=\mu_u+\mu_d+\mu_s= \mu_u+2\mu_d$ in the quark phase.

\begin{figure}
\centering
\includegraphics[width=12 cm]{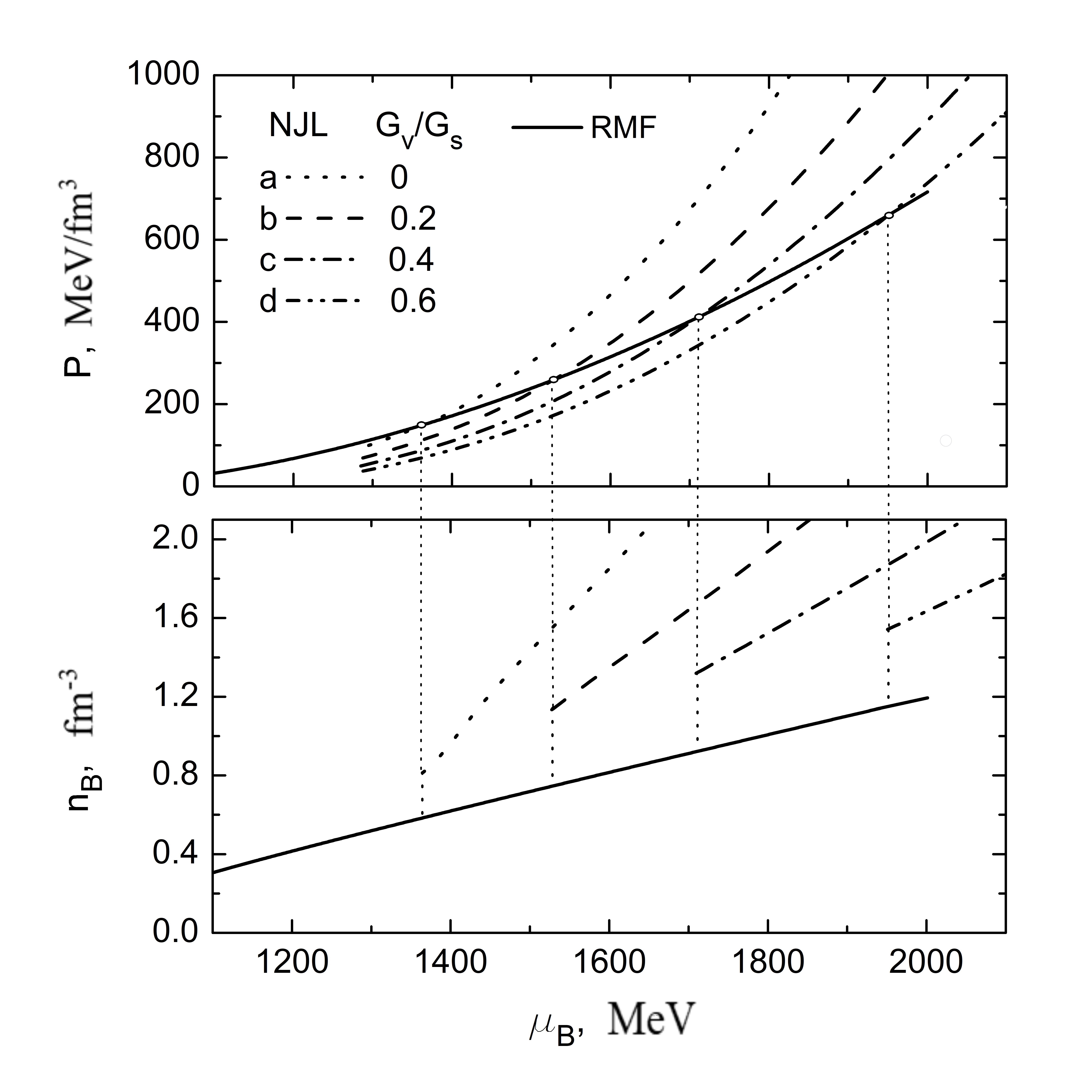}
\caption{\textbf{Top frame}: The pressure $P$ as a function of the baryonic
chemical potential $\mu_B$ for the hadronic matter in terms of the
RMF model (smooth curve) and for the deconfined quark matter
in the NJL model for different values of the vector coupling
constant: a) $G_V/G_S = 0$, b) $G_V /G_S = 0.2$, c) $G_V /G_S = 0.4$,
d) $G_V /G_S = 0.6$. \textbf{Bottom frame}: The density of the baryonic
charge $n_B$ as a function of the baryonic chemical potential $\mu_B$
in terms of the same models as in the top frame.}
\end{figure}

The top frame of Fig. 1 shows the dependence of the pressure $P$ on the baryonic chemical potential $\mu_B$ for hadronic matter obtained in the RMF model and for the deconfined quark matter obtained in the NJL model for four
values of the vector coupling constant $G_V=0$,  $G_V=0.2\,G_S$, $G_V=0.4\,G_S$, and $G_V=0.6\,G_S$. The figure shows that large values of the vector coupling constant for the interaction between the quarks correspond to a “stiffer” equation of state for the quark phase. We note that stiffer equations of state correspond to higher baryonic chemical potentials at equal pressure.

The intersection point of the curves for the hadronic and quark phases in the $P-\mu_B$ plane correspond to a state of equilibrium coexistence of the two phases. It can be seen that increasing the intensity of the vector-type
interaction between the quarks leads to an increase in the pressure of the phase coexistence $P_0$ and the baryonic chemical potential $\mu_{B0}$ in the state of an equilibrium coexistence of the two phases. We note that this kind of behavior is observed as well in the modified MIT bag model with a vector interaction \cite{39}. The bottom frame of Fig. 1 shows plots of the dependence of the baryonic charge $n_B$ on the baryonic chemical potential $\mu_B$ for the cases shown in the top frame of the figure. It is clear that in the state where the phases coexist, which corresponds to the intersection point of the curves in the $P-\mu_B$ plane, the densities of the baryonic charge for the hadronic phase and for the quark phase are different. At the same time, since the pressure $P$ and the baryonic chemical potential $\mu_B$ are continuously variable quantities, the baryonic charge density $n_B$ has a discontinuity (discontinuous change) at the phase transition point. The energy density has the same kind of behavior.

\begin{table}
\caption{First order phase transition parameters for different
ratios of vector and scalar coupling constants.} \centering
\begin{tabular}{cccccccc}
\toprule \textbf{Quark} & \textbf{$G_V/G_S$} & \textbf{$P_0$} &
\textbf{$\varepsilon_{H0}$} & \textbf{$n_{H0}$} &
\textbf{$\varepsilon_{Q0}$}
& \textbf{$n_{Q0}$} & \textbf{$\mu_{B0}$}  \\
 \textbf{model} &  & MeV/fm$^3$ &  MeV/fm$^3$ & fm$^{-3}$
& MeV/fm$^3$ & fm$^{-3}$ & MeV \\
\midrule
a & 0   & 150.2 &  646.88 & 0.5841 &  958.56 & 0.8128 & 1364.4  \\
b & 0.2 & 258.5 &  879.39 & 0.7449 & 1347.69 & 1.1343 & 1527.6  \\
c & 0.4 & 409.9 & 1163.47 & 0.9205 & 1515.03 & 1.3189 & 1709.4  \\
d & 0.6 & 659.5 & 1582.41 & 1.1495 & 1680.02 & 1.5420 & 1950.8  \\
\bottomrule
\end{tabular}
\end{table}

Table 1 shows the parameters of the phase transition of the first-order between the hadronic matter and the quark matter obtained by numerical solution of the equations of the coexistence of the two phases (29) for different values of the vector coupling constant of the quarks. Here $P_0$ is the pressure of the coexisting phases, $\mu_{B0}$ is the baryonic chemical potential, and $n_{H0}$ and $n_{Q0}$ are the densities of the baryonic charge of the hadronic and quark phases, respectively, while $\varepsilon_{H0}$ and $\varepsilon_{Q0}$ are the energy densities of the hadronic phase and quark phase, respectively, in the state where the phases coexist.

When the parameters of the phase transition and the equation of state of both the hadronic phase and the quark phase are known, it is possible to construct a hybrid equation of state of the superdense strongly interacting matter with a quark transition.

Figure 2 shows plots of the hybrid equations of state according to a Maxwell construction for a compact star with different values of the vector coupling constant for the interaction among the quarks.

\begin{figure}
\centering
\includegraphics[width=12 cm]{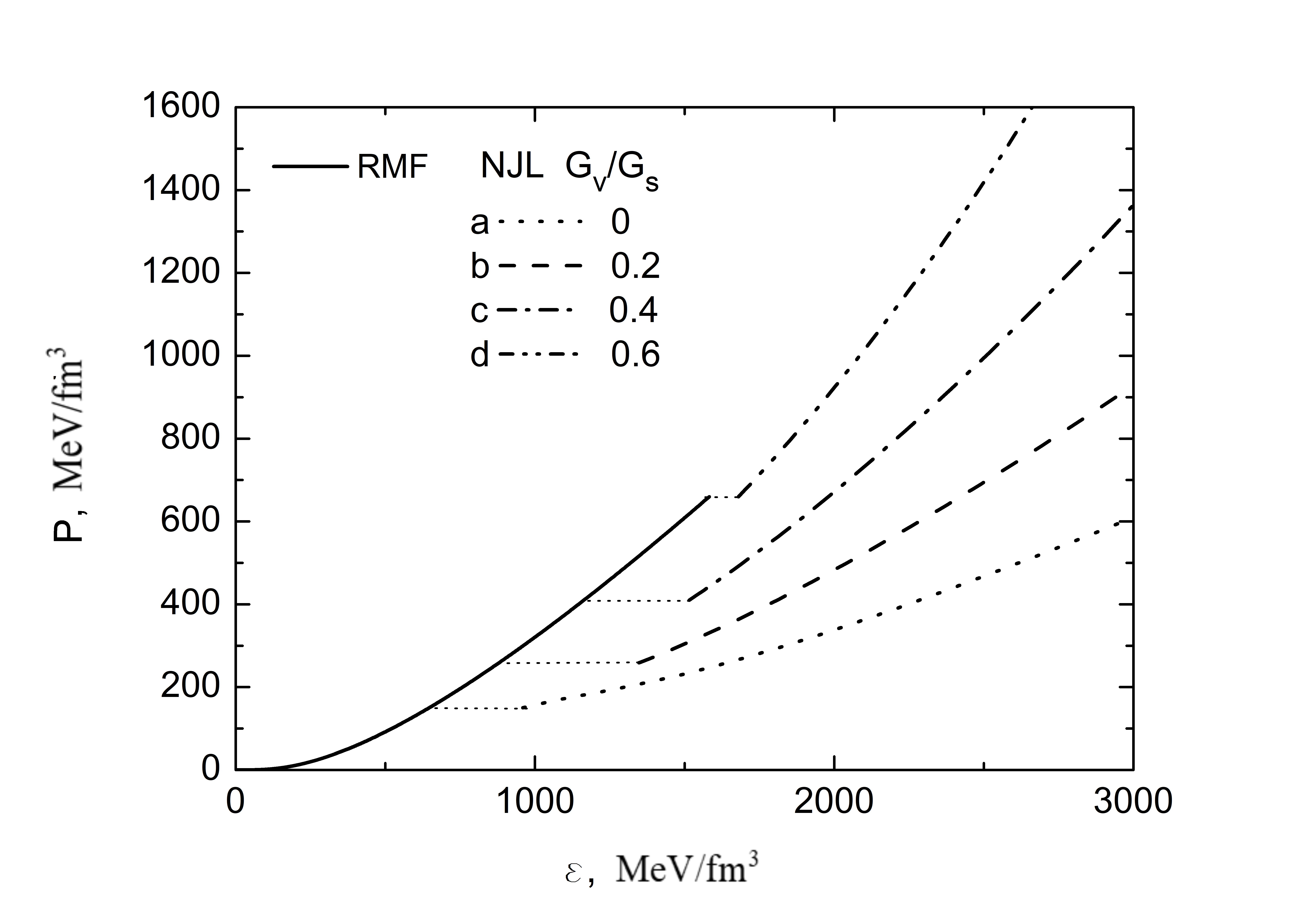}
\caption{Hybrid equations of state of matter in a compact
star for different values of the vector interaction coupling
constant of the quarks: a) $G_V /G_S = 0$, b) $G_V /G_S = 0.2$,
c) $G_V /G_S = 0.4$, d) G$_V /G_S = 0.6$.}
\end{figure}

If the phase transition for the deconfinement follows a Maxwell construction, then a compact star, with a pressure at its center which exceeds the value for the coexistence of the phases $P_0$ will have a central core of quark matter surrounded by ordinary matter with a hadronic structure. It is known that for a stable neutron star a decisive role is played by the sign of the derivative $dM/\varepsilon_c$. For $dM/\varepsilon_c>0$ the star is stable; otherwise it is unstable. An ordinary phase transition with constant pressure leads to a break in the plot of the mass of the neutron star on the central pressure. Depending on the value of the parameter of the jump, $\lambda=\varepsilon_{Q0}/(\varepsilon_{H0}+P_0)$, this break may lead to a change in the sign of the derivative $dM/\varepsilon_c$. For $\lambda\leq3/2$, the derivative $dM/\varepsilon_c$ does not change sign, while for $\lambda>3/2$ it does. This means that the newly formed infinitely small new phase will be stable for $\lambda\leq3/2$ and unstable for $\lambda>3/2$ \cite{40}.

For the hybrid equations of state obtained in this paper, the jump parameter $\lambda$ has the following values: a) $\lambda=1.20$ , b) $\lambda=1.18$, c) $\lambda=0.98$, d) $\lambda=0.75$. Thus, the hybrid equations of state with a quark phase transition that we have obtained ensure stability of the configuration of a compact star with a newly formed infinitely small core of quark matter.

\section{The structure of hybrid stars}

The hydrostatic equilibrium properties of spherically-symmetric and isotropic compact stars in the general theory of relativity are described by the Tolman-Oppenheimer-Volkoff (TOV) equations \cite{41,42}:
\begin{equation}
\label{eq30} 
\frac{dP}{dr}=-G\frac{(P+\varepsilon)\left( m+4\,\pi r^3P\right)}{r(r-2\,Gm)},~~~~\frac{dm}{dr}=4\,\pi r^2\varepsilon\,,
\end{equation}
where $G$ is the gravitational constant, $r$ is the distance from the star’s center, m is the so-called accumulated mass,
i.e., the mass inside a sphere of radius $r$, $P$ is the pressure, and $\varepsilon$ is the energy density at distance r from the star’s center. This system of equations becomes closed if the function $\varepsilon(P)$, i.e., the equation of state of the stellar matter, is known.

Numerical integration for a given value of the central pressure $P(0) = P_c$ begins at the center, where the
boundary condition $m(0) = 0$ is satisfied. The star’s radius $R$ is determined from the condition $P(R) = 0$, and the gravitational mass of the star is $M = m (r = R)$. 

Using the hybrid equations of state obtained in this paper, we have numerically integrated the system of TOV equations and determined such characteristics of compact stars as the mass $M$, radius $R$, and the profiles of the pressure and energy density $P(r)$ and $\varepsilon(r)$ respectively, for different values of the central pressure $P_c > P_0$. For configurations
with a central pressure $P_c > P_0$, the radius and mass of the quark core, $R_{core}$ and $M_{core}$, have also been determined.

In the range of densities corresponding to the inner and outer crust of the star, the well-known Baym-Bethe-Pethick equation of state \cite{43} was used.

\begin{figure}
\centering
\includegraphics[width=12 cm]{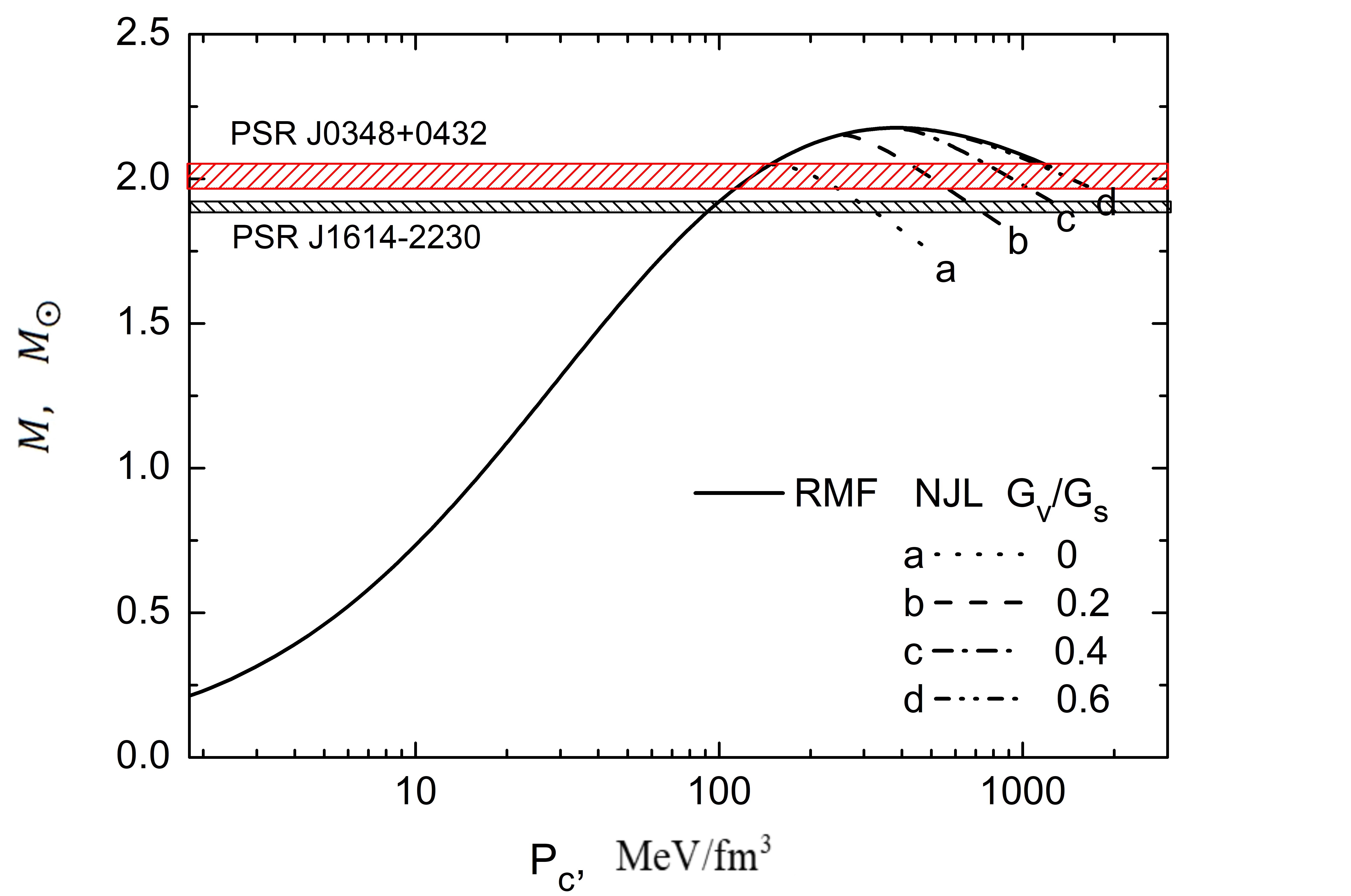}
\caption{The dependence of the gravitational mass of a
compact star on the central pressure for various values of the
vector coupling constant: a) $G_V/G_S = 0$, b) $G_V /G_S = 0.2$, c) $G_V /G_S = 0.4$,
d) $G_V /G_S = 0.6$. The smooth curve corresponds to
compact stars of hadronic matter, described in terms of the
RMF model. The horizontal shaded bands indicate
observational data from the pulsars PSR J0348+0432 and
PSR J1614-2230.}
\end{figure}

Figure 3 shows the dependence of the gravitational mass of a compact star on the central pressure for different values of the vector coupling constant: a) $G_V/G_S=0$ , b) $G_V/G_S=0.2$ , c) $G_V/G_S=0.4$ , d) $G_V/G_S=0.6$. The smooth curve corresponds to a compact star consisting of hadronic matter. Hybrid stars correspond to the branches denoted by a, b, c, and d, in accordance with the values of the vector coupling constant introduced above. In this figure the horizontal bands represent the results of measurements of the mass of the pulsar PSR J0348+0432 -
$M=(2.01\pm 0.04) M_\odot$ \cite{2} and refined data on the mass of PSR J1614-2230 $M=(1.908 \pm 0.016) M_\odot$ \cite{1,44,45}.

Since the jump parameter $\lambda$ for the hybrid equations of state we have examined satisfies the condition $\lambda<3/2$, for values of the central pressure slightly greater than the coexistence pressure $P_0$ of the phases, the derivative $dM/d\varepsilon_c$  satisfies the stability condition $dM/d\varepsilon_c>0$. As can be seen from Fig. 3, a fairly narrow, perhaps even unresolvable, range of values of the central pressure corresponds to stable hybrid stars. The maximum mass of hybrid stars is very close to the maximum mass of a hadronic star. A comparison of the results of our calculations with
observations of massive pulsars does not eliminate the possibility of deconfinement of quarks in the depths of a compact star.

Hybrid stars with a significant size of the quark core satisfy the condition of statistical instability $dM/d\varepsilon_c<0$. The condition for dynamic stability of a cold spherically-symmetric compact star found in a chemical equilibrium is the nonnegativity of the square of the fundamental mode of the small oscillations $( {\omega_0}^2\ge 0)$. Using a variational
principle, Chandrasekhar obtained an equation which makes it possible to determine numerically the eigenvalues of the different oscillatory modes [46] and clarify the stability question. When obtaining the equation it was assumed that the star consists of one phase and that the energy density inside the star varies continuously. Precisely for a continuous energy density inside the star it was shown that at the extremum point for the function $M(\varepsilon_c)$ the square of the fundamental mode ${\omega_0}^2$ changes sign. The Chandrasekhar equation is not applicable to the case of a compact star with a quark core, since the oscillations of a hybrid star, with a sharp boundary between two phases, will be accompanied by a process of mutual transformation of one phase into the other. The stability will depend on the relationship between the characteristic times of the oscillation and transformation of one phase into the other \cite{35,47,48,49}. It was shown \cite{35,47,48,49} that for a slow mutual conversion of the phases among stars on the branch of the $M(\varepsilon_c)$ curve, there will be some for which the condition $dM/d\varepsilon_c<0$ is satisfied, but these configurations will be stable with respect to the small radial oscillations. Hybrid stars of this type were referred to as slow-stable stars.

Figure 4 shows the mass-radius relationship for compact stars with different values of the vector-coupling
constant $G_V$ listed in Table 1. Compact stars consisting of hadronic matter correspond to the smooth curve, and the hybrid stars, to branches a, b, c, and d, in accordance with above values of the vector-coupling constant. The horizontal bands, as in Fig. 3, show the measured mass of the pulsar PSR J0348+0432 \cite{2} and a refined value of the mass of the pulsar PSR J1614-2230 \cite{44,45}, while the shaded rectangles indicate the regions corresponding to the
results of these observations of the mass and radius of the pulsars PSR J0030+0451 \cite{3} and PSR J0740+6620 \cite{5}.

\begin{figure}
\centering
\includegraphics[width=12 cm]{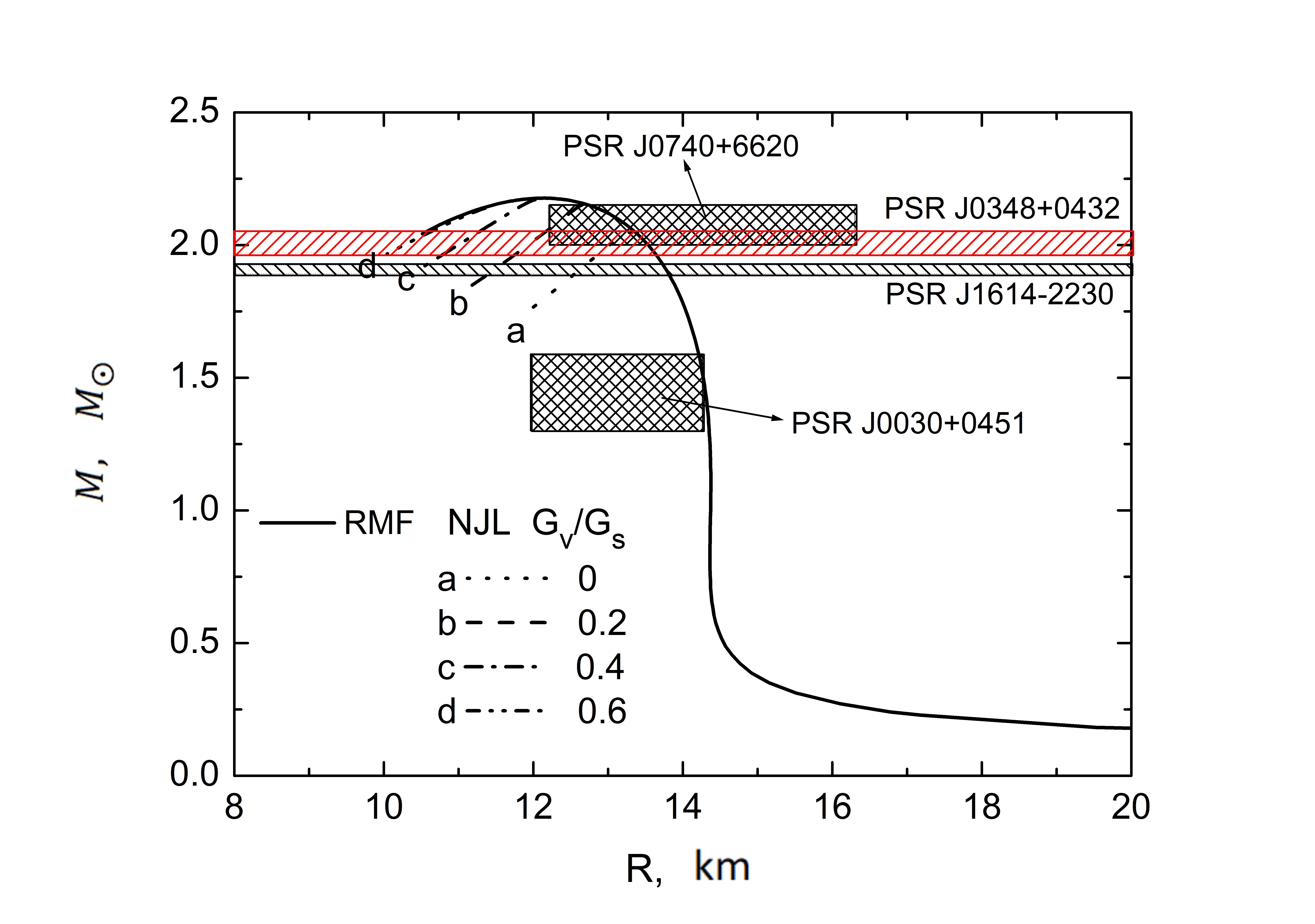}
\caption{ The mass-radius relation for a compact star with
different values of the vector-coupling constant: a) $G_V/G_S = 0$, b) $G_V /G_S = 0.2$, c) $G_V /G_S = 0.4$, d) $G_V /G_S = 0.6$. The smooth curve corresponds to compact stars of hadronic matter
described in terms of the RMF model. The horizontal
shaded strips indicate data from observations of the masses
of the pulsars PSR J0348+0432 and PSR J1614-2230. The
shaded rectangle on the bottom shows the results of
measuring the mass and radius of the pulsar PSR
J0030+0451 and the top rectangle, the results of analogous
measurements for PSR J0740+6620. }
\end{figure}

As Fig. 4 shows, increasing the intensity of the vector interaction increases the maximum mass of the compact
star and simultaneously reduces the radius of the star with the maximum mass. The theoretical calculations in the version of the RMF model that we have used yield an elevated value for the radius of the neutron star compared to that found by analyzing data from observations for the pulsar PSR J0030+0451 given in Ref. 3. Our results are close to the maximum values of the observations for the mass and radius of this pulsar. The results of our calculation are in agreement with the measured mass and radius of the pulsar PSR J0740+6620 \cite{5}.

\section{Conclusion}

In this paper a local SU(3) NJL model has been used to study the influence of the vector interaction channel
on the equation of state of quark matter, on the parameters of the phase transition between hadrons and quark matter, and on the structural characteristics of hybrid stars. The equation of state of the quark phase was found using values of the NJL model obtained in Ref. 23 by fitting from a reproduction of known experimental data on the characteristics of mesons. The only free parameter not subject to determination by this procedure in the model we have used is the vector coupling constant $G_V$. For the numerical calculations we have used four values of this constant: $G_V/G_S=0;~ 0.2;~ 0.4;~ 0.6$. Our calculations showed that for $G_V/G_S> 0.6$ the pressure of the phase transition $P_0$ exceeds the value of the pressure in the center of the compact star with the maximum mass with a hadronic structure.

To describe the hadronic phase we have used an expanded RMF model in which, besides the fields $\sigma$, $\omega$, and $\rho$ mesons, the field of the scalar-isovector $\delta$-meson was also taken into account. For the numerical calculations we used the values of the parameters of the RMF model obtained in Ref. 11.

Given that the phase transition from hadronic matter into quark matter is an ordinary phase transition of the
first-order described by a Maxwell construction, we determined the parameters of the phase transition for different values of the constant $G_V$.

It has been shown that high values of the vector coupling correspond of “stiffer” equations of state for the quark phase. Here when $G_V$ is higher the phase transition pressure $P_0$ is higher.

Using the obtained hybrid equations of state, the TOV equations have been integrated numerically and for
different values of the central pressure $P_c$ the mass and radius have been determined both for a compact star with a hadronic structure, and for a hybrid star with a quark core. It has been shown that when the vector constant is higher, the maximum mass of the compact star will be higher and the radius of the configuration with the maximum mass will be smaller.

A comparison of the results of the theoretical calculations with the measured masses of the currently known most massive pulsar PSR J0740+6620 shows that the resulting hybrid equations of state do not conflict with a limitation on the mass of $M_{max} \ge 2.01\, M_\odot$. The results are also consistent with the measured radius of this pulsar, $R\in [12.2 \div 16.3]$ km.

It has been shown that the density jump parameter $\lambda=\varepsilon_{Q0}/(\varepsilon_{H0}+P_0)$ for all the values of $G_V$ that we have examined satisfies the condition $\lambda<3/2$. This means that at the point of the phase transition in the $M-\varepsilon_c$ plane
there is a break such that the sign of the derivative $dM/d\varepsilon_c$ does not change, i.e., $dM/d\varepsilon_c>0$ both for configurations
of the hadronic branch and for configurations of the hybrid branch in the immediate vicinity of the critical
configuration. A hybrid star with an infinitely small central core of quark matter is stable. Hybrid stars with a substantial quark core satisfy the condition $dM/d\varepsilon_c<0$. If the mutual phase conversions take place fast enough, then these configurations are clearly unstable. In the case of a slow conversion, obtained neglecting the effects of phase transitions at the interface boundary of the phases during the time of small oscillations of a hybrid star, the instability condition $dM/d\varepsilon_c<0$ appears to be inapplicable (see, e.g., Ref. 35). In this case we are dealing with a slowly stable configuration.

\section*{Acknowledgments}
This work was carried out in the scientific-research laboratory for the physics of superdense stars in the department of applied electrodynamics and modeling at Yerevan State University, financed by the committee on science of the Ministry of Education, Science, Culture, and Sport of the Republic of Armenia.


\begin{thebibliography}{}
\bibitem{1}	P. Demorest, T.Pennucci, S.M. Ransom, et al., {\em Nature} {\bf 467}, 1081, 2010.
\bibitem{2}  J. Antoniadis, P.C.C. Freire, N. Wex, et al., {\em Science} {\bf 340}, 6131, 2013.
\bibitem{3}	M. Miller, F.K. Lamb, A. Dittmann, et al., {\em  Astrophys. J. Lett.} {\bf887}, L24, 2019.
\bibitem{4}	E.Fonseca, H.T.Cromartie, T.T.Pennucci, et al., 
{\em Astrophys. J. Lett.} {\bf915}, L12, 2021.
\bibitem{5}	M.C.Miller, F.K.Lamb, A.J.Dittmann, et al., 
{\em Astrophys. J. Lett.} {\bf918}, L28, 2021.
\bibitem{6}	K.Schertler, C.Greiner, J.Schaffner-Bielich, M.H.Thoma, {\em Nucl. Phys. A} {\bf677}, 463, 2000.
\bibitem{7}	G.F.Burgio, M.Baldo, P.K.Sahu, H.-J.Schulze, 
{\em Phys. Rev. C} {\bf66}, 025802, 2002.
\bibitem{8}	G.B.Alaverdyan, A,R.Harutyunyan, Yu.L.Vartanyan, 
{\em Astrophysics} {\bf46}, 361, 2003. 
\bibitem{9}	G.B.Alaverdyan, A,R.Harutyunyan, Yu.L.Vartanyan, 
{\em Astrophysics} {\bf47}, 52, 2004.
\bibitem{10}	B.K.Sharma, P.K.Panda, S.K.Patra, {\em Phys. Rev. C} {\bf75}, 035808, 2007.
\bibitem{11}	G.B.Alaverdyan, {\em Astrophysics} {\bf52}, 132, 2009.
\bibitem{12}	G.B.Alaverdyan, {\em Gravitation and Cosmology} {\bf15}, 5, 2009.
\bibitem{13}	A.G.Alaverdyan, G.B.Alaverdyan, A.O.Chiladze, 
{\em Int. J. Mod. Phys. D} {\bf19}, 1557, 2010.
\bibitem{14}	G.B.Alaverdyan, {\em Res. Astron. Astrophys} {\bf10}, 1255, 2010.
\bibitem{15}  R.Negreiros, V.A.Dexheimer, S.Schramm, {\em Phys. Rev. C} {\bf85}, 035805, 2012.
\bibitem{16}	G.B.Alaverdyan, Yu.L.Vartanyan, {\em Astrophysics} {\bf60}, 563, 2017. 
\bibitem{17}	S.Khanmohamadi, H.R.Moshfegh, S.Atashbar Tehrani, {\em Phys. Rev. D} {\bf101}, 023004, 2020.
\bibitem{18}	A.Chodos R.L.Jaffe, K.Johnson, et al., {\em Phys. Rev. D } {\bf9}, 3471, 1974.
\bibitem{19}	Y.Nambu, G.Jona-Lasinio, {\em Phys. Rev.} {\bf122}, 345, 1961. 
\bibitem{20}	Y.Nambu, G.Jona-Lasinio, {\em Phys. Rev.} {\bf124}, {\bf246}, 1961. 
\bibitem{21}	U.Vogl, W.Weise, {\em Prog. Part. Nucl. Phys.}{\bf 27}, 195, 1991.
\bibitem{22}	T.Hatsuda, T.Kunihiro, {\em Phys. Rep.} {\bf247}, 221, 1994.
\bibitem{23}	P.Rehberg, S.P.Klevansky, J.Hüfner, {\em Phys. Rev. C} {\bf53}, 410, 1996.
\bibitem{24}	M.Buballa, {\em Phys. Rep.} {\bf407}, 205, 2005.
\bibitem{25}	M.K.Volkov, A.E.Radzhabov, {\em Physics-Uspekhi}, {\bf176}, 569, 2006.
\bibitem{26}	P.Wang, A.W.Thomas, A.G.Williams, {\em Phys. Rev. C} {\bf75}, 045202, 2007.
\bibitem{27}	M.Alford, A.Sedrakian, {\em Phys. Rev. Lett.} {\bf119}, 161104, 2017.
\bibitem{28}    I.F.Ranea-Sandoval, M.G.Orsaria, G.Malfatti, et al., {\em Symmetry} {\bf11}, 425, 2019.
\bibitem{29}	J.J.Li, A.Sedrakian, M.Alford, {\em Phys. Rev. D} {\bf101}, 063022, 2020.
\bibitem{30}	H.Pais, D.P.Menezes, C.Providência, {\em Phys. Rev. C} {\bf93}, 065805, 2016.
\bibitem{31}	G.B.Alaverdyan, Yu.L.Vartanyan, {\em Astrophysics} {\bf61}, 483, 2018.
\bibitem{32}	J.D.Walecka, {\em Ann. Phys.} {\bf83}, 491, 1974.
\bibitem{33}	B.D.Serot, J.D.Walecka, {\em Int. J. Mod. Phys. E} {\bf6}, 515, 1997.
\bibitem{34}	G.B.Alaverdyan, {\em Symmetry} {\bf13}, 124, 2021. 
\bibitem{35}	G.Lugones, A.G.Grunfeld, {\em Universe} {\bf7}, 493, 2021.
\bibitem{36}	J.Boguta, A.R.Bodmer, {\em Nuclear Physics A} {\bf292}, 413, 1977.
\bibitem{37}	G.’t Hooft, {\em Phys. Rev. Lett.} {\bf37}, 8, 1976.
\bibitem{38}	N.K.Glendenning, {\em Phys. Rev. D} {\bf46}, 1274, 1992.
\bibitem{39}	M.Ju, J.Hu, H.Shen, {\em Astrophys. J.} {\bf923}, 250, 2021.
\bibitem{40}	Z.F.Seidov, {\em  Astron. Zh.} {\bf15}, 347, 1971.
\bibitem{41}	R.C.Tolman, {\em Phys. Rev.} {\bf55}, 364, 1939.
\bibitem{42}	J.R.Oppenheimer, G.M.Volkoff, {\em Phys. Rev.} {\bf55}, 374, 1939.
\bibitem{43}	G.Baym, H.Bethe, Ch.Pethick, {\em Nucl. Phys. A} {\bf175}, 255, 1971.
\bibitem{44}	E.Fonseca, T.T.Pennucci, J.A.Ellis, et al., {\em Astrophys. J.} {\bf832}, 167, 2016. 
\bibitem{45}	Z.Arzoumanian, A.Brazier, S.Burke-Spolaor, et al., {\em Astrophys. J. Suppl. Ser.} {\bf235}, 37, 2018.
\bibitem{46}  S.Chandrasekhar, {\em Astrophys. J.} {\bf140}, 417, 1964; Erratum in {\em Astrophys. J.} {\bf140}, 1342, 1964.
\bibitem{47}	J.P.Pereira, C.V.Flores, G.Lugones, 
{\em Astrophys. J.} {\bf860}, 12, 2018.
\bibitem{48}	L.Tonetto, G.Lugones, {\em Phys. Rev. D} {\bf101}, 123029, 2020.
\bibitem{49}	J.P.Pereira, M.Bejger, L.Tonetto, et al., 
{\em Astrophys. J.} {\bf910}, 145, 2021.
\end{thebibliography}
\end{document}